\def\year{2021}\relax
\documentclass[letterpaper]{article} 
\usepackage{aaai21}  
\usepackage{times}  
\usepackage{helvet} 
\usepackage{courier}  
\usepackage[hyphens]{url}  
\usepackage{graphicx} 
\usepackage{natbib}  
\usepackage{caption} 
\usepackage{booktabs} 

\usepackage[english]{babel}
\usepackage{moresize}
\usepackage{amsmath}

\usepackage{algorithmic} 
\usepackage{balance}
\usepackage{comment}
\usepackage{paralist}
\usepackage{multirow}

\usepackage{filecontents}

\usepackage[english]{babel}
\usepackage[latin1]{inputenc}
\usepackage{mathrsfs}
\usepackage{graphicx}
\usepackage{amssymb}
\usepackage{url}
\usepackage{subfigure}
\usepackage{amsmath}
\usepackage{enumitem}
\usepackage[linesnumbered,algoruled,boxed,lined]{algorithm2e}
\usepackage{adjustbox}
\usepackage{amssymb}
\usepackage{times}
\usepackage{hyperref}

\usepackage{pgfplots}
\usetikzlibrary{pgfplots.dateplot}

\usepackage{filecontents}
\definecolor{tblue}{RGB}{31,119,180}
\definecolor{torange}{RGB}{255,127,14}
\definecolor{tgreen}{RGB}{44,160,44}
\definecolor{tred}{RGB}{214,39,40}
\definecolor{tpurple}{RGB}{148,103,189}

\usepackage{filecontents}

\newcommand{\hide}[1]{} 

\newcommand{\ie}{\textit{i}.\textit{e}.}
\newcommand{\eg}{\textit{e}.\textit{g}.} 
\newcommand{\wrt}{\textit{w}.\textit{r}.\textit{t}}

\title{Knowledge-Enhanced Hierarchical Graph Transformer Network \\for Multi-Behavior Recommendation}

\author{
Lianghao Xia$^1$, Chao Huang$^2$\thanks{Corresponding author: Chao Huang}, Yong Xu$^{1,3,4}$, Peng Dai$^2$, Xiyue Zhang$^1$ \\\Large{\bf Hongsheng Yang$^2$, Jian Pei$^5$, Liefeng Bo$^2$}\\
}
\affiliations{
    \textsuperscript{\rm 1}South China University of Technology, China, \textsuperscript{\rm 2}JD Finance America Corporation, USA\\
    \textsuperscript{\rm 3}Communication and Computer Network Laboratory of Guangdong, China\\
    \textsuperscript{\rm 4}Peng Cheng Laboratory, Shenzhen, China, \textsuperscript{\rm 5}Simon Fraser University, Canada\\
\{cslianghao.xia,zhang.xiyue\}@mail.scut.edu.cn, chaohuang75@gmail.com, yxu@scut.edu.cn\\
\{peng.dai,hongsheng.yang,liefeng.bo\}@jd.com, jpei@cs.sfu.ca

}

\def\model{KHGT}

\begin{document}

\maketitle

\begin{abstract}
Accurate user and item embedding learning is crucial for modern recommender systems. However, most existing recommendation techniques have thus far focused on modeling users' preferences over singular type of user-item interactions. Many practical recommendation scenarios involve multi-typed user interactive behaviors (\eg, page view, add-to-favorite and purchase), which presents unique challenges that cannot be handled by current recommendation solutions. In particular: i) complex inter-dependencies across different types of user behaviors; ii) the incorporation of knowledge-aware item relations into the multi-behavior recommendation framework; iii) dynamic characteristics of multi-typed user-item interactions. To tackle these challenges, this work proposes a \underline{\textbf{K}}nowledge-Enhanced \underline{\textbf{H}}ierarchical \underline{\textbf{G}}raph \underline{\textbf{T}}ransformer Network (\model), to investigate multi-typed interactive patterns between users and items in recommender systems. Specifically, \model\ is built upon a graph-structured neural architecture to i) capture type-specific behavior characteristics; ii) explicitly discriminate which types of user-item interactions are more important in assisting the forecasting task on the target behavior. Additionally, we further integrate the graph attention layer with the temporal encoding strategy, to empower the learned embeddings be reflective of both dedicated multiplex user-item and item-item relations, as well as the underlying interaction dynamics. Extensive experiments conducted on three real-world datasets show that \model\ consistently outperforms many state-of-the-art recommendation methods across various evaluation settings. Our implementation code is available in \emph{https://github.com/akaxlh/KHGT}.
\end{abstract}

\section{Introduction}
\label{sec:intro}

Recommender systems have been widely deployed in many Internet services (\eg, e-commerce, online review and advertising systems) to alleviate information overload and deliver the most relevant items to users~\cite{liu2020diversified,2019online}. In the recommendation scenario with the focus on implicit feedback, Collaborative Filtering (CF) becomes one of most popular paradigm which factorizes user-item interactions into latent representations and predicts user's preference based on the projected low-dimensional embeddings~\cite{chen2020efficient}.

Many deep neural network techniques have been developed to enhance collaborative filtering architecture for non-linear feature interactions. Specifically, early studies, like NCF~\cite{he2017neural} and DMF~\cite{xue2017deep} utilize the Multi-layer Perceptron to handle the non-linear interactions. Furthermore, autoencoder-based methods are designed for mapping high-dimensional sparse user-item interactions into low-dimensional dense representations~\cite{sedhain2015autorec,wu2016collaborative}. Later works investigate the use of graph neural network to exploit the high-order user-item relations, and perform neighborhood-based feature aggregations~\cite{zhang2019star,wang2019neural}.

Although these methods have shown promising results, a deficiency is that they only model singular type of user-item interactions, which makes them insufficient to distill the complex collaborative signals from the multi-typed behaviors of users~\cite{mbgcn2020}. In particular, there typically exist multiple relations between user and item that exhibit various behavior characteristics in many real-world recommendation scenarios, which are particularly helpful in learning users' preferences on the target type of behavior~\cite{guo2019buying}. For example, in online retail platforms, users' page view and add-to-favorite activities over different items, can serve as the auxiliary knowledge for assisting the forecasting task of customer purchase (target behavior). Therefore, it is crucial to take such inter-type behavioral influences into consideration to more accurately infer user preferences.

There are several key technical challenges that remain to be solved to realize the multi-behavior recommendation. \emph{First}, how to distill the user-specific collaborative signals from the multiplex user-item interaction behaviors, is a significant challenge to tackle. In practice, type-specific behavioral patterns interweave with each other in a complex manner~\cite{gao2019learning}, like the complementary correlations between the add-to-cart and purchase behaviors, or users' negative reviews are mutually exclusive with their positive feedback over the same item. Without the explicitly encoding of such heterogeneous relationships between user and item, models may suffer from the inability of capturing the complicated inherent cross-type behavior dependencies in a hierarchical way. \emph{Second}, another core challenge lies in the incorporation of knowledge-aware item semantic relatedness into the encoding function of multi-behavioral patterns. The knowledge-aware side information often contains much fruitful facts and contextual connections about items~\cite{wang2019knowledge}. It is desirable to rigorously design a joint embedding paradigm over the user-item and item-item relations in our multi-behavior recommendation. \emph{Third}, a time-aware model is needed to better handle the temporal information of user-item interactions.

There are a handful of recent models that attempt to integrate multi-behavioral interactive patterns for making recommendations~\cite{mbgcn2020,gao2019learning}. However, these works intend to consider multi-typed interactions in a relatively independent and local manner (\eg, singular dimensional cascading correlations), and can hardly capture the high-order multiplex relationships across users and items. Additionally, how to account for the side knowledge from items as well as user-item interaction dynamics, is less explored in those multi-behavior recommender systems.

In light of these differences and challenges, we present a general framework--\underline{\textbf{K}}nowledge-Enhanced \underline{\textbf{H}}ierarchical \underline{\textbf{G}}raph \underline{\textbf{T}}ransformer Network (\model), for multi-behavior recommendation. Particularly, at the first stage, we develop a multi-behavior graph transformer network which performs recursive embedding propagation, to capture behavior heterogeneity across users and items in an attentive aggregation schema. As a result, user-item and item-item relationships of different types are enabled to maintain their specific representation space. To handle behavior dynamics, we inject the time-aware context into the graph transformer framework through a temporal encoding strategy. In addition, to encode the inter-dependencies between type-specific behavior representations, a multi-behavior mutual attention encoder is proposed to learn dependent structures of different types of behaviors in a pairwise manner. Finally, a gated aggregation layer is introduced to discriminate the contribution of type-specific relation embeddings for making recommendations.





The contributions of this paper are highlighted as follows:

\begin{itemize}[leftmargin=*]

\item We propose a framework \model, which explicitly achieves high-order relation learning in the knowledge-aware multi-behavior collaborative graph under the hierarchically structured graph transformer network.

\item To jointly integrate user- and item-wise collaborative similarities under the multi-behavior modeling paradigm of \model: i) the first-stage graph-structured transformer module captures the type-specific user-item interactive patterns in a time-aware environment; ii) the second-stage attentive fusion network encodes the cross-type behavior hierarchical dependencies and discriminates the type-specific contribution, in forecasting the target behaviors.

\item We apply the proposed \model\ method to three real-world datasets of movie, venue and product recommendations. Experiments show that our model achieves significant gains over many state-of-the-art baselines from various lines. Furthermore, model interpretation ability is also investigated with case studies of qualitative examples.


\end{itemize}




\section{Preliminaries}
\label{sec:model}

We begin with the introduction of key notations and consider a typical recommendation scenario with $I$ users ( $U=\{u_1,...,u_i,...,u_I\}$) and $J$ items ($V=\{v_1,...,v_j,...,v_J\}$). We further define the relevant graph-structured data as: \\\vspace{-0.12in}

\noindent \textbf{User-Item Multi-Behavior Interaction Graph}. $G_u$. With the awareness of different types of user-item interactions, a multi-behavior interaction graph is defined as: $G_u=(U,V,E_u)$, in which the edge set $E_u$ represents $K$ types of relations (\eg, browse, add-to-favorite, purchase) between user and item. In $E_u$, each edge $e_{i,j}^k$ denotes that user $u_i$ interacts with item $v_j$ under the behavior type of $k$.\\\vspace{-0.12in}

\noindent \textbf{Knowledge-aware Item-Item Relation Graph}. $G_v$. To incorporate the side information of items, we define graph $G_v=(V,E_v)$ to characterize multiplex dependencies across items with the consideration of their external knowledge. In $G_v$, edge $e_{j,j'}^r$ linked between item $v_j$ and $v_{j'}$ with their meta relations, which is denoted as $\{(v_j,r,v_{j'}) | v_j,v_{j'}\in V, r\in R\}$. Here, $R$ indicates the set of relations which can be generated from different aspects, such as $v_j$ and $v_{j'}$ belong to the same category, from the same location, or interacted with the same user under the same behavior type.\\\vspace{-0.12in}




\noindent \textbf{Task Formulation}. Based on above definitions, we formulate the knowledge-aware multi-behavior recommendation:

\textbf{Input}: the user-item multi-behavior interaction graph $G_u$ and the knowledge-aware item-item relation graph $G_v$.


\textbf{Output}: a predictive model which effectively infers the probability $y_{i,j}$ of unseen interaction between user $u_i$ and item $v_j$ under the target behavior type of $k$.
\section{Methodology}
\label{sec:solution}

We elaborate the details of \model, which investigates the multiplex user-item relations in an end-to-end manner.\vspace{-0.1in}

\subsection{Attentive Heterogeneous Message Aggregation}\vspace{-0.05in}
In this component, we aim to collectively capture the multi-behavior user-item interactive patterns and item-item dependencies in a unified graph-structured neural network. The overall architecture is shown in Figure~\ref{fig:framework}. \vspace{-0.05in}

\subsubsection{Multi-Behavior Interactive Pattern Encoding.}
This module aims to aggregate the heterogeneous signals from multi-behavioral patterns between the user and his/her interacted items. Towards this end, we develop an adaptive multi-behavior self-attention network under a message passing paradigm with the enhancement of global behavior context, which consists of three key modules: temporal context encoding scheme, information propagation and aggregation.\\\vspace{-0.12in}

\begin{figure}
	\centering
	\includegraphics[width=0.48\textwidth]{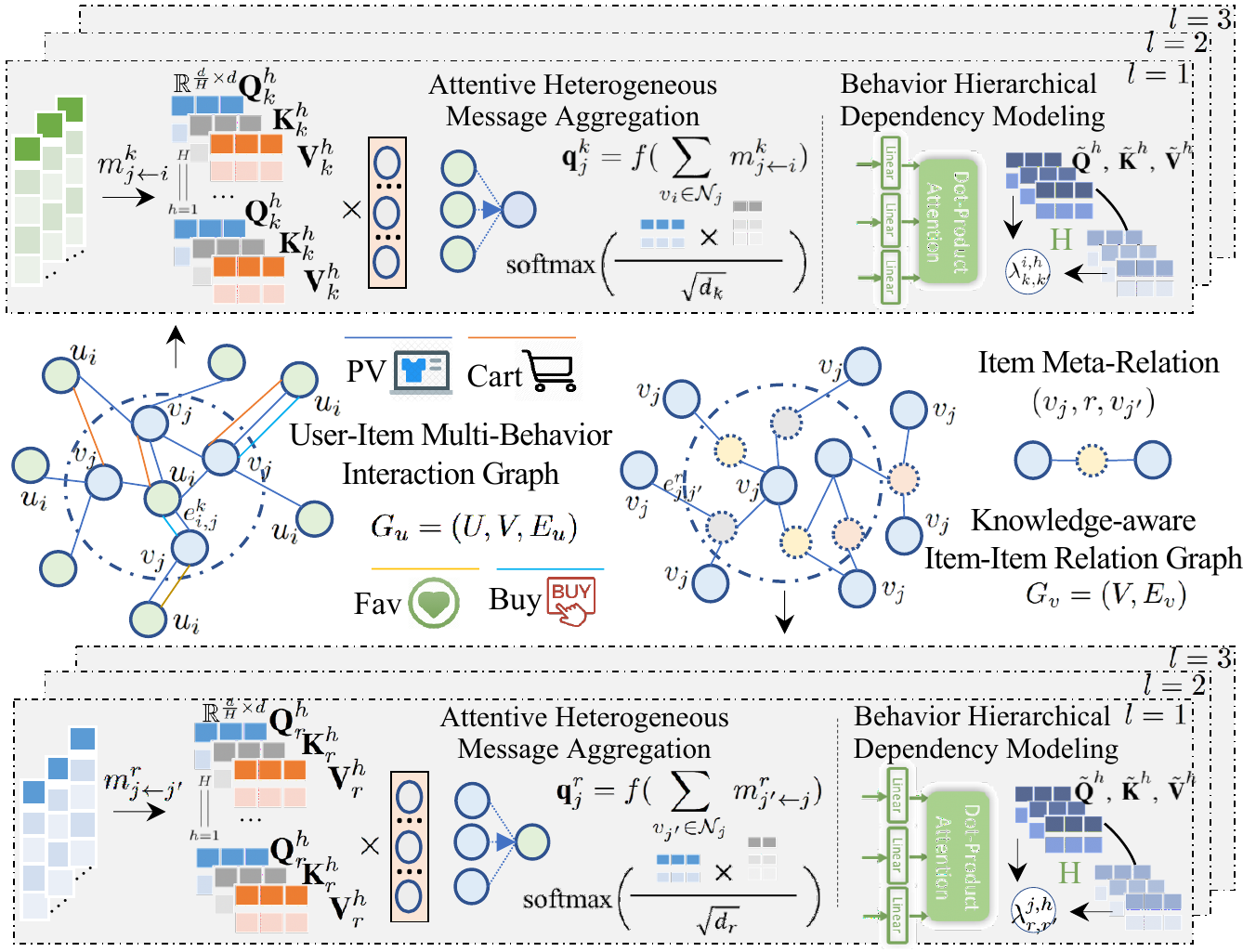}
	\vspace{-0.3in}
	\caption{The model architecture of \model\ framework.}
	\label{fig:framework}
	\vspace{-0.2in}
\end{figure}

\noindent \textbf{Temporal Information Encoding.}
To capture the influences between different types of user-item interactions in a time-aware scenario, we develop a temporal context encoding scheme, to incorporate the behavior dynamics in our heterogeneous message aggregation architecture. In particular, given the connection $E_{i,j}^k$ between user $u_i$ and item $v_j$ under the behavior of $k$, we map their corresponding interaction timestamp $t_{i,j}^k$ into the time slot as $\tau(t_{i,j}^k)$, and utilize the sinusoidal functions for embedding $\textbf{T}_{i,j}^k \in \mathbb{R}^{2d}$ generation, which is motivated by the positional embedding framework in Transformer~\cite{vaswani2017attention,hu2020heterogeneous,2020hierarchical}.
\begin{small}
\begin{align}
\textbf{T}_{(i,j)}^{k,(2l)} = sin(\frac{\tau(t_{i,j}^k)}{10000^{\frac{2l}{d}}}), \textbf{T}_{(i,j)}^{k,(2l+1)} = cos(\frac{\tau(t_{i,j}^k)}{10000^{\frac{2l+1}{d}}}) 
\end{align}
\end{small}
\noindent where the element indexs (even and odd position index) in the temporal information embedding are represented as $(2l)$ and $(2l+1)$, respectively. $d$ is the latent dimensionality. To augment the tunable ability of our temporal context encoding, we further apply a projection layer over $\textbf{T}_{i,j}^k$ as: $\bar{\textbf{T}}_{i,j}^k = \textbf{T}_{i,j}^k \cdot \textbf{W}_{k}$, $\textbf{W}_k \in \mathbb{R}^{2d\times d}$. $\textbf{W}_k$ is the transformation weights corresponding to $k$-th type of interactions. \\\vspace{-0.1in}

\noindent \textbf{Information Propagation Phase.}
We perform the time-aware information propagation between the source and target node, over the multi-behavior user-item interaction graph $G_u$, with the following graph attentive mechanism:
\begin{align}
m_{i\leftarrow j}^k = \mathop{\Bigm|\Bigm|}\limits_{h=1}^{H} \omega_{i,j,k}^h \cdot \textbf{V}_k^h \textbf{p}_j;m_{j\leftarrow i}^k = \mathop{\Bigm|\Bigm|}\limits_{h=1}^{H} \omega_{j,i,k}^h \cdot \textbf{V}_k^h \textbf{p}_i
\end{align}
\noindent where $m_{i\leftarrow j}^k$ and $m_{j\leftarrow i}^k$ denote the propagated message from item $v_j$ to user $u_i$, and from $u_i$ to $v_j$, respectively. $\textbf{p}_j$ is the element-wise addition between initialized item embedding $\textbf{e}_{j}$ and the corresponding temporal context representation $\bar{\textbf{T}}_{i,j}^k$. \ie, $\textbf{p}_j= \textbf{e}_j \oplus \bar{\textbf{T}}_{i,j}^k$. Similar operation is applied for the message from user side: $\textbf{p}_i= \textbf{e}_i \oplus \bar{\textbf{T}}_{i,j}^k$. $\textbf{V}^h_k\in\mathbb{R}^{\frac{d}{H}\times d}$ is the $h$-head projection matrix with respect to the $k$-th behavior type. In addition, $\omega_{i,j,k}^h$, $\omega_{j,i,k}^h$ represent the learned attentive propagation weights over the constructed message $\textbf{p}_j$ and $\textbf{p}_i$, respectively. Formally, they are calculated as:
\begin{small}
\begin{align}
\bar{\omega}_{i,j,k}^h = \frac{(\textbf{Q}_k^h \textbf{p}_i)^\top (\textbf{K}_k^h \textbf{p}_j)}{\sqrt{d/H}};~
\omega_{i,j,k}^h=\frac{\exp(\bar{\omega}_{i,j,k}^h)}{\Sigma_{e_{i,j'}^k\in E_u}\exp(\bar{\omega}_{i,j',k}^h)} \nonumber
\end{align}
\end{small}
\noindent where $\textbf{Q}_k^h, \textbf{K}_k^h \in \mathbb{R}^{\frac{d}{H}\times d}$ are the head-specific query and key transformation with respect to the $k$-th behavior type.



To incorporate the global context across different behavior types in the message passing process, we learn the attention-based transformation matrices $\textbf{Q}_k^h, \textbf{K}_k^h, \textbf{V}_k^h$ in a multi-channel parameter learning framework. To be specific, we design a base transformation paradigm which consists of $M$ channels of parameters, \ie, $\bar{\textbf{Q}}_m^h, \bar{\textbf{K}}_m^h, \bar{\textbf{V}}_m^h$ ($m=1...M$). They correspond to $M$ latent projection subspaces, which reflect different aspects of the common behavior context across different types. Formally, the type-specific transformation procedure is performed with the gating mechanism:
\begin{small}
\begin{align}
\textbf{Q}_k^h=\sum_{m=1}^M \alpha_{m}^k \bar{\textbf{Q}}_m^h, \textbf{K}_k^h=\sum_{m=1}^M \beta_{m}^k \bar{\textbf{K}}_m^h, \textbf{V}_k^h=\sum_{m=1}^M \gamma_{m}^k \bar{\textbf{V}}_m^h
\end{align}
\end{small}
where $\alpha_{m}^k, \beta_{m}^k, \gamma_{m}^k$ are quantitative gated weights for the $m$-th channel transformation with respect to behavior type of $k$. Typically, the number of channels $M$ is smaller than the number of behavior types $K$ in practice, which enables the parameter-efficient heterogeneous message aggregation.\\\vspace{-0.12in}

\noindent \textbf{Information Aggregation Phase.} Based on the constructed propagated message $m_{i\leftarrow j}^k$ and $m_{j\leftarrow i}^k$, we aggregate the neighboring information with the summation operation:
\begin{small}
\begin{align}
\textbf{q}_i^k=f(\sum_{v_j\in\mathcal{N}_i} m_{i\leftarrow j}^k);~~~~
\textbf{q}_j^k=f(\sum_{v_i\in\mathcal{N}_j} m_{j\leftarrow i}^k)
\end{align}
\end{small}
where $\mathcal{N}_i$ and $\mathcal{N}_j$ denote the neighborhood of $u_i$ and $v_j$ in the user-item interaction graph $G_u$. $\textbf{q}_i^k, \textbf{q}_j^k$ are the aggregated information of $u_i$ and $v_j$ under the behavior type of $k$. $f(\cdot)$ is an activation function like LeakyReLU.

\subsubsection{Information Aggregation for Item-side Relations.}
We fuse the heterogeneous signals from the item-item inter-dependencies with the attentive aggregation:
\begin{small}
\begin{align}
\textbf{q}_j^r = f(\sum_{v_{j'}\in\mathcal{N}_j}m_{j\leftarrow j'}^r) = \sum_{v_{j'}\in\mathcal{N}_{j}} \mathop{\Bigm|\Bigm|}\limits_{h=1}^{H} \omega_{j,j',r}^h \textbf{V}_r^h \textbf{p}_{j'}
\end{align}
\end{small}
\noindent where $r$ is the index of $R$ different item-item relations and $\mathcal{N}_{j}$ indicates the neighboring nodes of $v_j$ over the graph $G_v$. \vspace{-0.1in}

\begin{figure}[t]
	\centering
	\includegraphics[width=0.47\textwidth]{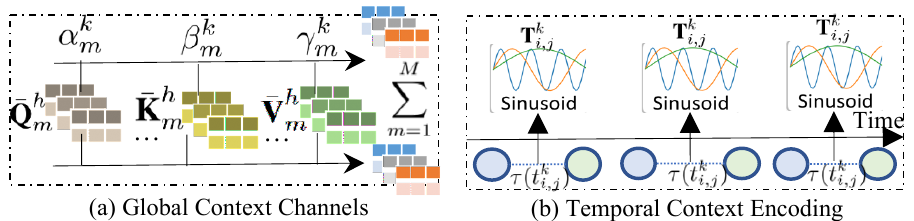}
	\vspace{-0.25in}
	\caption{The framework of global context enhanced parameter learning and behavior dynamics encoding in \model.}
	\label{fig:framework}
	\vspace{-0.25in}
\end{figure}

\subsection{Behavior Hierarchical Dependency Modeling}
In our multi-behavior recommendation scenarios, user behaviors with different types interact with each other in a complex and hierarchical way. To address this challenge, two questions arise: i) how do we effectively preserve mutual relations between different types of behavior; ii) how to promote the collaboration across different type-specific behavior representations in assisting the final recommendation.

Our mutual relation encoder is developed based on the scaled dot-product attention via learning the pairwise type-wise relevance scores $\lambda_{k,k'}^{i,h}$, which is formally represented:
\begin{small}
\begin{align}
\label{eq:multihead}
\tilde{\textbf{q}}_{i}^k &= \text{MH-Att}(\textbf{q}_{i}^k)=\mathop{\Bigm|\Bigm|}\limits_{h=1}^H \sum_{k'=1}^K  \lambda_{k,k'}^{i,h} \cdot \tilde{\textbf{V}}^{h} \cdot {\textbf{{q}}}_{i}^{k'} \\
\lambda_{k,k'}^{i,h} & = \frac{\exp{\bar{\lambda}}_{k,k'}^{i,h}}{\sum_{k'=1}^K \exp{\bar{\lambda}^{i,h}_{k,k'}}}; \bar{\lambda}_{k,k'}^{i,h}=\frac{(\tilde{\textbf{Q}}^h \cdot \textbf{q}_{i}^k)^\top (\tilde{\textbf{K}}^h \cdot \textbf{q}_{i}^{k'})}{\sqrt{d/H}} \nonumber
\end{align}
\end{small}
\noindent where $\tilde{\textbf{Q}}^h$, $\tilde{\textbf{K}}^h$, $\tilde{\textbf{V}}^h$ are learnable projection matrices of the $h$-th learning subspace. Similar operations are applied for updating the item embedding $\tilde{\textbf{q}}_{j}^k$.

Next, we propose to fuse the learned type-specific behavior representations, by investigating the individual importance in forecasting the target type of user interactions. We present our gated fusion mechanism for conclusive representation $\mathbf{\Phi}_j$ as:
\begin{align}
\mathbf{\Phi}_j= \sum_{k=1}^K \eta_{j}^k \tilde{\textbf{q}}_{j}^k \oplus \sum_{r=1}^R \xi_{j}^r \tilde{\textbf{q}}_{j}^r ; \eta_{j}^k= \sigma(\bar{\eta}_{j}^{k}); \xi_{j}^r= \sigma(\bar{\xi}_{j}^{r}) 
\end{align}
\noindent $\sigma(\cdot)$ is the softmax function. $\eta_{j}^k$ and $\xi_{j}^r$ are the learned importance score of $k$-th type of user-item interaction representation $\tilde{\textbf{q}}_{i}^k$, and $r$-th type of item-item relation representation $\tilde{\textbf{q}}_{j}^r$, respectively. They are formally calculated as:
\begin{align}
\bar{\eta}_{j}^k=\textbf{a}_0^\top f(\tilde{\textbf{q}}_{j}^k)+{c}_0^u;~~\bar{\eta}_{j}^r=\textbf{b}_0^\top f(\tilde{\textbf{q}}_{j}^r)+{c}_0^v
\end{align}


\noindent $f(\cdot)$ denotes the multi-layer network as: $f(\tilde{\textbf{q}}_{j}^k)=\textbf{B}_1^u \tilde{\textbf{q}}_{j}^k + \textbf{B}_2^u \sum_{k'=1}^K \tilde{\textbf{q}}_{j}^k+\textbf{c}_1^u$. Parameters in our gated fusion layer are denoted as $\textbf{a}_0$, $\textbf{b}_0$, $\textbf{B}_1^*$, $\textbf{B}_2^*$ (transformations), $\textbf{c}_0^*$ and $\textbf{c}_1^*$ (bias terms). This gating mechanism is also utilized for obtaining user representation $\mathbf{\Phi}_i$ over type-specific embeddings $\tilde{\textbf{q}}_{i}^k$.\\\vspace{-0.1in}


\noindent \textbf{High-order Multi-Behavior Pattern Propagation.}
Based on the defined information propagation and aggregation functions, we capture high-order collaborative relations under the multi-behavior context (user-item interaction graph $G_u$) in our graph neural network. The update procedure from the $(l)$-th layer to the $(l+1)$-th layer is ($\mathbf{\Phi}_j^{(l)} \in \mathbb{R}^d$):
\begin{align}
\mathbf{\Phi}_j^{(l+1)} \leftarrow \mathop{\textbf{Aggregate}}\limits_{i\in N_u(j); j'\in N_v(j)} \Big ( \textbf{Propagate}( \mathbf{\Phi}_{i}^l, \mathbf{\Phi}_{j'}^l, G) \Big )
\end{align}
\noindent Propagate($\cdot$) is the information propagation function which extracts useful features from both the user-item interactions (over $E_u$) and item-item dependencies (over $E_v$). Aggregate($\cdot$) denotes the information fusion function. The final embeddings are summarized cross different order-specific representations as: $\mathbf{\Phi}_j = \mathbf{\Phi}_j^{(1)} \oplus ... \oplus \mathbf{\Phi}_j^{(L)}$.

\subsection{The Learning Phase of \model}
After generating the conclusive representations $\mathbf{\Phi}_i$ and $\mathbf{\Phi}_j$ for users and items, the likelihood of $u_i$ interacting with $v_j$ under the target behavior can be inferred as $\text{Pr}_{i,j} = \textbf{z}^\top \cdot (\mathbf{\Phi}_i \odot \mathbf{\Phi}_j)$. To perform the model optimization, we aim to minimize the following marginal pair-wise loss function:
\begin{align}
    \mathcal{L}=\sum_{i=1}^I\sum_{s=1}^S \max(0,1-\text{Pr}_{i,p_s}+\text{Pr}_{i,n_s})+\lambda\|\mathbf{\Theta}\|_\text{F}^2
\end{align}
where $I$ denotes the number of trained users, and $S$ denotes the number of positive-negative pairs for each user. In practice, we randomly sample $S$ positive items $v_{p_1}, v_{p_2},...,v_{p_s}$ and $S$ negative items $v_{n_1}, v_{n_2},...,v_{n_s}$ for each user. In addition, $\mathbf{\Theta}$ represents the set of all trainable parameters, and $\lambda$ is the weight for the regularization term.\vspace{-0.05in}

\subsubsection{Sub-graph Sampling for Large-Scale Data.}
One key challenge of graph neural architecture lies in the information aggregation over the entire graph in a full-batch mode, which consumes tremendous memory and computation cost. To endow \model\ with the ability of handling large-scale data, we develop a random-walk-based sub-graph sampling algorithm over the graph $G_u$ and $G_v$, and maintain a weight vector during the sampling process based on node relatedness extracted from the adjacent matrix of the graph.


\subsubsection{Model Complexity Analysis.}
Our \model\ spends $O(K\times(I+J)\times d^2)$ to calculate the $\textbf{Q}, \textbf{K}, \textbf{V}$ transformations, and $O(|E|\times d)$ for information aggregation. For type-wise relation modeling, the most prominent computation comes from the $O(K\times (I+J)\times d^2)$ transformations. Overall, our \model\ could achieve comparable time complexity with the GNN-based multi-behavior recommendation methods. Also, \model\ costs moderate extra memory for the intermediate results, compared to the most existing GNN models.


\vspace{-0.01in}
\section{Evaluation}
\label{sec:eval}

\vspace{-0.02in}
This section answers the following research questions:\vspace{-0.05in}
\begin{itemize}
\item \textbf{RQ1}: How does \emph{\model} perform compared with the various state-of-the-art recommender systems?\vspace{-0.05in}
\item \textbf{RQ2}: How do different designed modules and captured relational structures contribute to the model performance?\vspace{-0.05in}
\item \textbf{RQ3}: How  does \emph{\model} work with the integration of different types of behavior in our heterogeneous aggregator?\vspace{-0.05in}
\item \textbf{RQ4}: How does \emph{\model} perform \wrt\ different interaction sparsity levels as compared to representative competitors?\vspace{-0.05in}
\item \textbf{RQ5}: How does \emph{\model} perform with different parameter settings (\eg, latent dimensionality and GNN depth)?\vspace{-0.05in}
\item \textbf{RQ6}: How is the interpretation ability of our \emph{\model} in capturing behavior inter-dependencies?
\end{itemize}
\vspace{-0.07in}

\subsection{Experimental Settings}

\begin{table}[t]
    \centering
    \scriptsize
	\setlength{\tabcolsep}{0.6mm}
    \begin{tabular}{ccccc}
        \toprule
        Dataset&User \#&Item \#&Interaction \#&Interactive Behavior Type\\
        \midrule
        Yelp&19800&22734&$1.4\times 10^6$&\{Tip, Dislike, Neutral, Like\}\\
        ML10M&67788&8704&$9.9\times 10^6$&\{Dislike, Neutral, Like\}\\
        Online Retail&805506&584050&$6.4\times 10^7$&\{Page View, Favorite, Cart, Purchase\}\\
        \bottomrule
    \end{tabular}
    \caption{Statistics of the experimented datasets}
    \label{tab:data}
\vspace{-0.15in}
\end{table}

\subsubsection{\bf Data Description.}
\label{sec:data}
We show the data statistics in Table~\ref{tab:data}.\\\vspace{-0.12in}

\noindent\textbf{MovieLens Data}.
We differentiate explicit user's rating scores $r$ (\ie, $[1,...,5]$) into multiple behavior types: (1) $r\leq 2$: \emph{dislike} behavior. (2) 2$<r<$4. \emph{neutral} behavior. (3) $r \geq 4$: \emph{like} behavior. In this data, the target and auxiliary behaviors are set as (like) and (neutral, dislike), respectively.

\noindent\textbf{Yelp Data}.
User's feedbacck interactions over items in this data, are projected into three behavior types by following the same partition rubric of MovieLens. We regard the like behavior as the target type and (dislike, neutral, tip) as auxiliary sources, where \emph{tip} behavior indicates that users gave tips on their visited venues.

\noindent \textbf{Online Retail Data}.
We also investigate our \emph{\model} in a real-world online retail scenario with explicit multi-typed user-item interactions, \ie, \emph{page view}, \emph{add-to-cart}, \emph{add-to-favorite} and \emph{purchase}. For this application, the target behavior is set as purchases and the other three types of user behaviors are considered as auxiliary behavioral signals.

For the above applications, the knowledge-aware item-item relation graph $G_v$ is generated with item meta-relations from two dimensions: i) $v_j$--$u_i$--$v_{j'}$ under the behavior type of $k$; ii) the categorical relations between item $v_j$ and $v_{j'}$.

\subsubsection{\bf Evaluation Protocols.}
We adopt two metrics: \textit{Normalized Discounted Cumulative Gain (NDCG@$N$)} and \textit{Hit Ratio (HR@$N$)} which have been widely used in recommendation tasks~\cite{wang2019neural,chen2018sequential}. Note that higher HR and NDCG scores signal better performance. Following the same settings in~\cite{yu2019multi,zhao2020adversarial}, we employ the time-aware leave-one-out evaluation to split the training/test sets. The test set contains the last interactive item of users and the rest of data is used for training. For fair and efficient evaluation, each positive instance is paired with randomly selected 99 non-interactive items, which is consistent with the experimental settings in~\cite{sun2019bert4rec,huang2019taxonomy}.

\subsubsection{\bf Methods for Comparison.} Baselines are presented as:
\label{sec:baseline}

\noindent \textbf{Conventional Matrix Factorization Approach}:
\begin{itemize}[leftmargin=*]
\item \textbf{BiasMF}~\cite{koren2009matrix}: it enhances the matrix factorization paradigm by incorporating user and item bias with the corresponding implicit feedback.
\end{itemize}

\noindent \textbf{Autoencoder-based Collaborative Filtering}:
\begin{itemize}[leftmargin=*]
\item \textbf{AutoRec}~\cite{sedhain2015autorec}: it is a stacked autoencoder architecture to project user-item interactions into hidden representation unit with the data reconstruction loss.
\item \textbf{CDAE}~\cite{wu2016collaborative}: it designs a denoising Autoencoder for user-item interaction modeling.
\end{itemize}

\noindent \textbf{Neural Network-enhanced Collaborative Filtering}:
\begin{itemize}[leftmargin=*]
\item \textbf{DMF}~\cite{xue2017deep}: it enhances the MF with a multi-layer perceptron to encode the interaction vector of users.
\item \textbf{NCF}~\cite{he2017neural}: Three variants of NCF with different interaction encoders: \ie, Multilayer perceptron (NCF-M), concatenated element-wise-product branch (NCF-N) and the fixed element-wise product (NCF-G).
\end{itemize}

\noindent \textbf{Neural Auto-regressive Recommendation Methods}:
\begin{itemize}[leftmargin=*]
\item \textbf{NADE}~\cite{zheng2016neural}: it incorporates the parameter sharing mechanism into the autoregressive CF model.
\item \textbf{CF-UIcA}~\cite{du2018collaborative}: it is a neural CF framework with auto-regression on user-item correlations. 
\end{itemize}

\noindent \textbf{Graph Neural Networks Collaborative Filtering}:
\begin{itemize}[leftmargin=*]
\item \textbf{ST-GCN}~\cite{zhang2019star}: it is a graph-structured encoder-decoder framework to learn latent embeddings of users and items under data scarcity, via GCNs.
\item \textbf{NGCF}~\cite{wang2019neural}: it is a message passing architecture to exploit high-order connection relationships.
\end{itemize}

\noindent \textbf{Recommendation with Multi-Behavioral Patterns}.
\begin{itemize}[leftmargin=*]
\item \textbf{NMTR}~\cite{gao2019neural}: it is a multi-task recommendation framework which explores the cascaded correlations between multiple types of user-item interactive behavior.
\item \textbf{DIPN}~\cite{guo2019buying}: this approach jointly considers the behavior patterns of browsing and buying with the bi-directional recurrent network based attention mechanism.
\item \textbf{NGCF$_M$}~\cite{wang2019neural}: we integrate the multi-behavioral relation learning with the neural graph collaborative filtering model under a joint graph neural network.
\item \textbf{MATN}~\cite{xia2020multiplex}: it learns the type-wise interaction dependencies with a memory attention network.
\item \textbf{MBGCN}~\cite{mbgcn2020}: it models the multi-behavior of users and uses graph convolutional network to perform behavior-aware embedding propagation.
\end{itemize}

\noindent \textbf{Knowledge-aware Recommendation Method}.
\begin{itemize}[leftmargin=*]
\item \textbf{KGAT}~\cite{wang2019kgat}: it investigates the high-order connectivity of the semantic item relations over the collaborative knowledge graph, with GAT framework.
\end{itemize}

\subsubsection{\bf Parameter Settings.}
We implement \emph{\model} with TensorFlow and infer the model parameters with Adam optimizer. Our multi-head self-attention module is configured with the 2 heads for embedding learning. The channels of base transformations in our graph attention module is set as 2. The model training process is performed with the learning rate of $1e^{-3}$ (with decay rate of 0.96) and batch size of 32. The regularization strategy with the weight decay, which is chosen from the set of \{0.1, 0.05, 0.01, 0.005, 0.001\}. This is applied in the training phase to alleviate the overfitting issue.

\begin{table*}[t]
\centering
\ssmall
\setlength{\tabcolsep}{1mm}
\begin{tabular}{|c|c|c|c|c|c|c|c|c|c|c|c|c|c|c|c|c|c|c|c|}
\hline
Data & Metric & BiasMF & ~DMF~ & NCF-M & NCF-G & NCF-N & AutoRec & ~CDAE~ & NADE & CF-UIcA & ST-GCN & NGCF & NMTR & DIPN & NGCF$_M$ & MBGCN & MATN & KGAT & \emph{\model}\\
\hline
\multirow{2}{*}{Yelp}
&HR & 0.755 & 0.756 & 0.714 & 0.755 & 0.771 & 0.765 & 0.750 & 0.792 & 0.750 & 0.775 & 0.789 & 0.790 & 0.791 & 0.793 & 0.796 & 0.826 & 0.835 & \textbf{0.880} \\
\cline{2-20}
\cline{2-20}
&NDCG & 0.481 & 0.485 & 0.429 & 0.487 & 0.500 & 0.472 & 0.462 & 0.499 & 0.469 & 0.465 & 0.500 & 0.478 & 0.500 & 0.492 & 0.502 & 0.530 & 0.543 & \textbf{0.603}\\
\cline{2-20}
\hline
\multirow{2}{*}{MovieLens}
&HR & 0.767 & 0.779 & 0.757 & 0.787 & 0.801 & 0.658 & 0.659 & 0.761 & 0.778 & 0.738 & 0.790 & 0.808 & 0.811 & 0.825 & 0.826 & 0.847 & 0.817 & \textbf{0.861}\\
\cline{2-20}
\cline{2-20}
&NDCG & 0.490 & 0.485 & 0.471 & 0.502 & 0.518 & 0.392 & 0.392 & 0.486 & 0.491 & 0.444 & 0.508 & 0.531 & 0.540 & 0.546 & 0.553 & 0.569 & 0.514 & \textbf{0.597} \\
\cline{2-20}
\hline
\multirow{2}{*}{Retail}
&HR & 0.262 & 0.305 & 0.319 & 0.290 & 0.325 & 0.313 & 0.329 & 0.317 & 0.332 & 0.347 & 0.302 & 0.332 & 0.317 & 0.374 & 0.369 & 0.354 & 0.377 & \textbf{0.464} \\
\cline{2-20}
\cline{2-20}
&NDCG & 0.153 & 0.189 & 0.191 & 0.167 & 0.201 & 0.190 & 0.196 & 0.191 & 0.198 & 0.206 & 0.185 & 0.179 & 0.178 & 0.221 & 0.222 & 0.209 & 0.214 & \textbf{0.278} \\
\cline{2-20}
\hline
\end{tabular}
\vspace{-0.12in}
\caption{Performance comparison on Yelp, MovieLens, Online Retail data, in terms of \textit{HR@$k$} and \textit{NDCG@$k$} ($k=10$).}
\vspace{-0.15in}
\label{tab:target_behavior}
\end{table*}

\subsection{Performance Validation (RQ1)}
We first present the performance of all methods in forecasting the target type of user-item interactions on three datasets in Table~\ref{tab:target_behavior}. From the results, we observe that remarkable performance improvement can be achieved by \emph{\model} on different types of datasets. Such performance gap can be attributed to the joint exploration of multi-type behavior inter-dependencies and the underlying knowledge-aware item collaborative signals.

Among various competitive methods, the recommender systems (\eg, NMTR, NGCF$_M$, MBGCN) which consider multi-typed interactions, improve the performance as compared to other baselines. This points to the positive effect of aggregating multiplex behavioral patterns in the designed interaction encoding function. Furthermore, GNN-based neural approaches perform better than autoencoder and autoregressive CF models, suggesting the rationality of exploring high-order collaborative signals over user-item relations. We investigate the ranking quality of top-$k$ items recommended by different methods ranging from 1 to 9. Table~\ref{tab:vary_k} lists the results of Yelp data. We can observe that the best performance is always achieved by \emph{\model} under different top-$N$ settings.

\begin{table}[h]
    \vspace{-0.05in}
	\centering
    \scriptsize
	\setlength{\tabcolsep}{0.6mm}
	\begin{tabular}{|c|c|c|c|c|c|c|c|c|c|c|}
		\hline
		\multirow{2}{*}{Model}&\multicolumn{2}{c|}{@1}&\multicolumn{2}{c|}{@3}&\multicolumn{2}{c|}{@5}&\multicolumn{2}{c|}{@7}&\multicolumn{2}{c|}{@9}\\
		\cline{2-11}
		&HR&NDCG&HR&NDCG&HR&NDCG&HR&NDCG&HR&NDCG\\
		\hline
		\hline
        NADE&0.265&0.265&0.508&0.402&0.642&0.454&0.720&0.478&0.784&0.497\\
        \hline
        CF-UIcA&0.235&0.235&0.449&0.360&0.576&0.412&0.659&0.440&0.731&0.463\\
        \hline
        ST-GCN&0.216&0.216&0.445&0.347&0.580&0.400&0.669&0.430&0.744&0.454\\
        \hline
        NMTR&0.214&0.214&0.466&0.360&0.610&0.419&0.700&0.450&0.762&0.469\\
        \hline
        MATN&0.279&0.279&0.529&0.423&0.659&0.477&0.741&0.507&0.798&0.524\\
        \hline
        KGAT&0.291&0.291&0.546&0.439&0.684&0.496&0.763&0.521&0.823&0.538\\
        \hline
        \emph{\model} & \textbf{0.355} & \textbf{0.355} & \textbf{0.617} & \textbf{0.506} & \textbf{0.748} & \textbf{0.559} & \textbf{0.818} & \textbf{0.583} & \textbf{0.864} & \textbf{0.599}\\
		\hline
	\end{tabular}
	\vspace{-0.1in}
	\caption{Ranking performance evaluation on Yelp dataset with varying Top-\textit{K} value in terms of \textit{HR@K} and \textit{NDCG@K}}
	\label{tab:vary_k}
	\vspace{-0.2in}
\end{table}


\vspace{-0.05in}
\subsection{Model Ablation Study (RQ2)}
\begin{figure}[h]
	\centering
	\vspace{-0.1 in}
	\vspace{-0.07 in}
	\subfigure[][Yelp]{
		\centering
		\includegraphics[width=0.3\columnwidth]{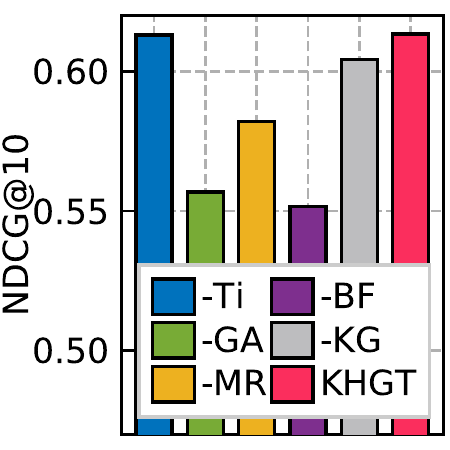}
		\label{fig:ab_yelp_NDCG}
	}
	\subfigure[][MovieLens]{
		\centering
		\includegraphics[width=0.3\columnwidth]{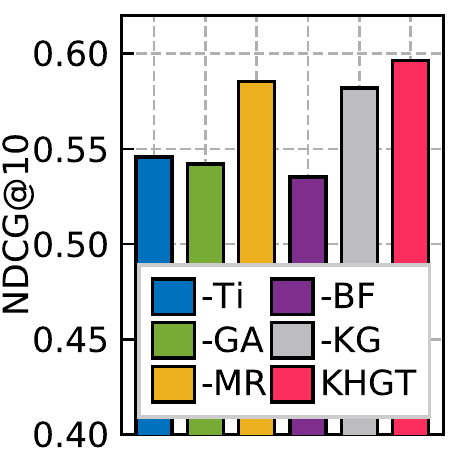}
		\label{fig:ab_ml10m_NDCG}
	}
	\subfigure[][Retail]{
		\centering
		\includegraphics[width=0.3\columnwidth]{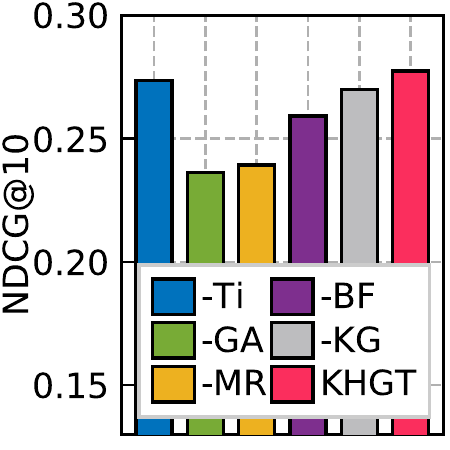}
		\label{fig:ab_retail_NDCG}
	}
	\vspace{-0.2in}
	\caption{Ablation studies of sub-modules in \model.}
	\label{fig:ablation}
\end{figure}

We consider different model variants of \emph{\model} from five perspectives and analyze their effects (as shown in Figure~\ref{fig:ablation}):

\noindent \textbf{Type-specific Behavioral Pattern Modeling}. \emph{\model}-GA. We first evaluate the effect of our type-specific behavior semantic learning by replacing our attentive heterogeneous information aggregation with graph convolutional network. \\\vspace{-0.12in}

\noindent \textbf{Behavior Mutual Dependency Modeling}. \emph{\model}-MR. We remove the transformer-based mutual relation encoder to capture inter-dependencies of different types of behaviors. \\\vspace{-0.12in}

\noindent \textbf{Cross-Type Behavioral Pattern Fusion}. \emph{\model}-BF. This simplified variant directly applies the mean pooling operation over the type-specific behavior representations, instead of using the designed gated fusion layer.\\\vspace{-0.12in}

\noindent \textbf{Temporal Context Encoding}. \emph{\model}-Ti. This variant does not include the temporal context embedding when performing information propagation in our graph transformers.\\\vspace{-0.12in}

\noindent \textbf{Incorporation of Item-Item Relations}. \emph{\model}-KG. It does not integrate the item-item relations in the framework.

We can observe that the full version of our developed \emph{\model} achieves the best performance in all cases. We further summarize the conclusions: (1) Modeling the type-specific user-item interactive patterns in an explicit attentive way, is better than performing graph-structured convolutions. (2) The efficacy of augmenting the multi-behavior recommendation with the underlying mutual relation learning. (3) The necessity of explicit discrimination for the contributions of type-specific behavior patterns. (4) The positive effect of temporal context information in capturing the behavior dynamics. (5) The incorporation of item external knowledge in our graph neural network is helpful for more accurately encoding user's multi-dimensional preference.

\subsection{Performance v.s. Multi-Behavior Integration (RQ3)}
\begin{figure}[t]
	\centering
	\vspace{-0.1 in}
	\subfigure[][Yelp-HR@\textit{k}]{
		\centering
		\includegraphics[width=0.22\textwidth]{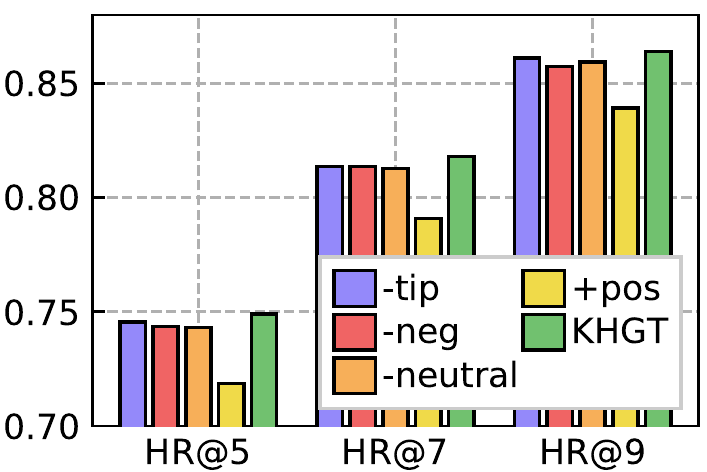}
		\label{fig:beh_yelp_hr}
	}
	\vspace{-0.1 in}
	\subfigure[][Retail-HR@\textit{k}]{
		\centering
		\includegraphics[width=0.22\textwidth]{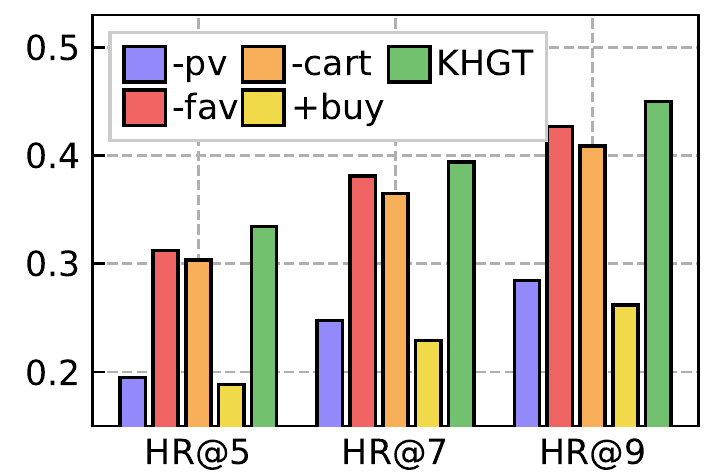}
		\label{fig:beh_onelineRetail_hr}
	}
	\vspace{-0.1in}
	\caption{Impact study of different behavior types on Yelp and Online Retail data, in terms of \textit{HR@k} and \textit{NDCG@k}.}
	\label{fig:behavior_compare}
	\vspace{-0.2in}
\end{figure}

We conduct the ablation experiments to validate whether the incorporation of more diverse behavior types could boost the performance. The model variants are generated with rubric as follows: (1) ``+''behavior type indicates only using the target behavior itself to make predictions (\eg, +like, +buy). (2) ``-''behavior type means removing this specific type of interactions (\eg, -pv: page-view, -cart: add-to-cart) in forecasting the user's target behavior. From the results in Figure~\ref{fig:behavior_compare}, \emph{\model} using all types of interaction behaviors consistently achieves the best performance compared to other variants, which suggests that more diverse behavior incorporation could improve the recommendation results with more comprehensive multi-behavior knowledge.

\subsection{Influences of Interaction Sparsity Degrees (RQ4)}
\begin{figure}
    \centering
	\subfigure[][Retail-HR@10]{
		\centering
		\includegraphics[width=0.22\textwidth]{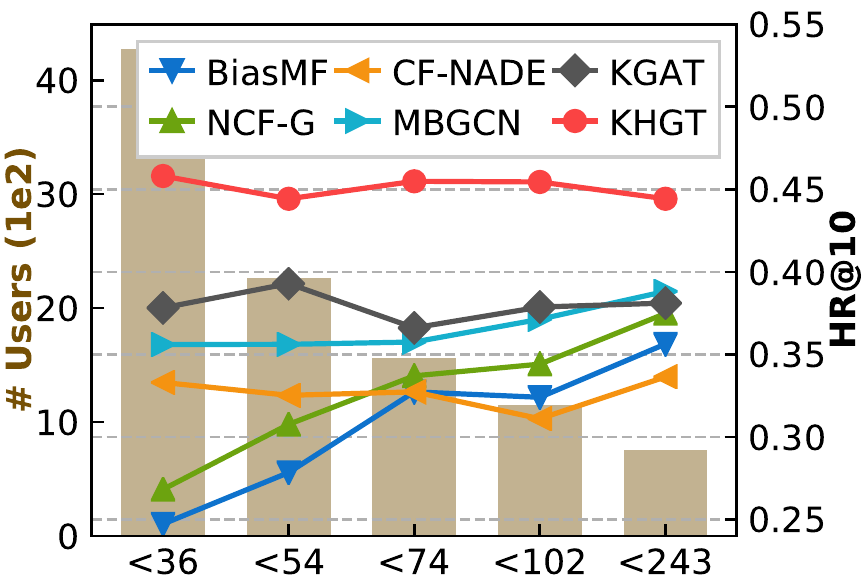}
		\label{fig:sp_eco_hr}
	}
	\subfigure[][Retail-NDCG@10]{
		\centering
		\includegraphics[width=0.22\textwidth]{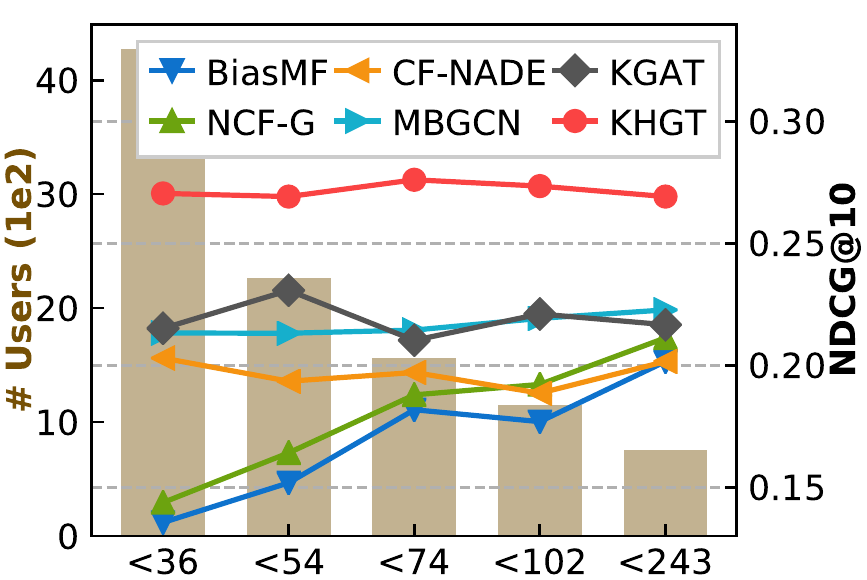}
		\label{fig:sp_eco_ndcg}
	}
	\vspace{-0.15in}
	\caption{Model performance on the online Retail data \textit{w.r.t.} different data sparsity, in terms of \textit{HR@10} and \textit{NDCG@10}.}
	\label{fig:sparsity}
	\vspace{-0.1in}
\end{figure}

we further perform experiments to evaluate the model performance with respect to different  sparsity levels of user-item interaction data. In particular, each user is grouped in one of five different data sparsity degrees in terms of his/her interaction density over items. The bars in the background of Figure~\ref{fig:sparsity} show the number of users which belong to the corresponding sparsity levels in x-axis. We keep the total number of interaction summation of each sparsity level as the same. The y-axis shows the recommendation accuracy of \emph{\model} and several representative baselines. We can notice that the performance gap between our approach and other competitors become more significant as data becomes more sparse, which also ascertains the reasonableness of \emph{\model} in enhancing recommender systems with the capability of learning complex interactive patterns, by modeling inter-dependencies among various types of user behaviors.

\subsection{Hyperparameter Effect Investigation (RQ5)}
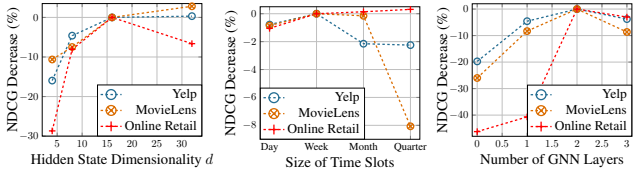
\begin{figure}
    \centering
    \begin{adjustbox}{max width=1.0\linewidth}

\begin{tikzpicture}
\begin{axis}[
    xlabel={Hidden State Dimensionality $d$},
    ylabel={NDCG Decrease (\%)},
    xmin=2,xmax=34,
    ymin=-30,ymax=3.5,
    legend columns=1,
    legend cell align=right,
    grid=both,
    every axis plot/.append style={ultra thick},
    every tick label/.append style={scale=1.3},
    label style={scale=1.8},
    legend style={
        nodes={scale=1.5, transform shape},
        legend image post style={scale=1.5},
        },
    legend style={at={(1,0)},anchor=south east},
    every axis plot post/.append style={
        every mark/.append style={scale=2}
    }
]
\addplot[color={rgb:blue,4;green,2;yellow,1}, mark=o, dashed, mark options={solid}]
table[x=para, y=yelp_ndcg] {latFactor-buy.txt};
\addplot[color={rgb:red,4;green,1;yellow,2}, mark=otimes, dashed, mark options={solid}]
table[x=para, y=ml10m_ndcg] {latFactor-buy.txt};
\addplot[color={rgb:red,4}, mark=+, dashed, mark options={solid}]
table[x=para, y=eco_ndcg] {latFactor-buy.txt};
\legend{\large Yelp, \large MovieLens, \large Online Retail};

\end{axis}
\end{tikzpicture}



\begin{tikzpicture}
\begin{axis}[
    xlabel={Size of Time Slots},
    ylabel={NDCG Decrease (\%)},
    xmin=0.8,xmax=4.2,
    ymin=-9,ymax=0.5,
    xtick={1,2,3,4},
    xticklabels={Day, Week, Month, Quarter},
    legend columns=1,
    legend cell align=right,
    grid=both,
    every axis plot/.append style={ultra thick},
    every tick label/.append style={scale=1.3},
    label style={scale=1.8},
    legend style={
        nodes={scale=1.5, transform shape},
        legend image post style={scale=1.5},
        },
    legend style={at={(0,0)},anchor=south west},
    every axis plot post/.append style={
        every mark/.append style={scale=2}
    }
]
\addplot[color={rgb:blue,4;green,2;yellow,1}, mark=o, dashed, mark options={solid}]
table[x=para, y=yelp_ndcg] {memoSize-buy.txt};
\addplot[color={rgb:red,4;green,1;yellow,2}, mark=otimes, dashed, mark options={solid}]
table[x=para, y=ml10m_ndcg] {memoSize-buy.txt};
\addplot[color={rgb:red,4}, mark=+, dashed, mark options={solid}]
table[x=para, y=eco_ndcg] {memoSize-buy.txt};
\legend{\large Yelp, \large MovieLens, \large Online Retail};
\end{axis}
\end{tikzpicture}


\begin{tikzpicture}
\begin{axis}[
    xlabel={Number of GNN Layers},
    ylabel={NDCG Decrease (\%)},
    xmin=-0.1,xmax=3.1,
    ymin=-48,ymax=2,
    legend columns=1,
    legend cell align=right,
    grid=both,
    every axis plot/.append style={ultra thick},
    every tick label/.append style={scale=1.3},
    label style={scale=1.8},
    legend style={
        nodes={scale=1.5, transform shape},
        legend image post style={scale=1.5},
        },
    legend style={at={(1,0)},anchor=south east},
    every axis plot post/.append style={
        every mark/.append style={scale=2}
    }
]
\addplot[color={rgb:blue,4;green,2;yellow,1}, mark=o, dashed, mark options={solid}]
table[x=para, y=yelp_ndcg] {gnnLayer-buy.txt};
\addplot[color={rgb:red,4;green,1;yellow,2}, mark=otimes, dashed, mark options={solid}]
table[x=para, y=ml10m_ndcg] {gnnLayer-buy.txt};
\addplot[color={rgb:red,4}, mark=+, dashed, mark options={solid}]
table[x=para, y=eco_ndcg] {gnnLayer-buy.txt};
\legend{\large Yelp, \large MovieLens, \large Online Retail};

\end{axis}
\end{tikzpicture}
    \end{adjustbox}
    \vspace{-0.20in}
    \caption{Hyper-parameter study for the like/purchase prediction in terms of \textit{HR@10} and \textit{NDCG@10}.}
    \vspace{-0.1in}
    \label{fig:hyperparam}
\end{figure}

We show the parameter study results of \emph{\model} in Figure~\ref{fig:hyperparam}.

\begin{itemize}[leftmargin=*]
\item \textbf{Embedding Dimensionality: $d$}. The model achieves better performance when we increase $d$ from 4 to 16, since a larger hidden dimension representation might be beneficial to capture more latent factors for user-item interactions. Due to the overfitting phenomenon, the performance degrades with the further increase of $d$. \vspace{-0.05in}
\item \textbf{Time Resolution of Temporal Encoding: $\tau(\cdot)$}. The best performance is achieved with the projected time slot of week resolution, which indicates that the weekly interactive patterns is a good trade-off between the modeling of user-specific behavior dynamics and interweave correlations with others.\vspace{-0.05in}
\item \textbf{Depth of Graph Neural Network: $L$}. By stacking more graph neural layers to jointly capture the high-order user-item and item-item collaborative relations, the recommendation performance is improved. \emph{\model} with two recursive message propagation layers achieves the best results.\vspace{-0.05in}
\item \textbf{Sub-graph Sampling Scale $N$}. Table~\ref{tab:graphSampling} shows the performance when varying the sampled sub-graph size (measured by \# of nodes). We observe that training with smaller sub-graph size (training with dropout regularization to alleviate overfitting), and relatively larger sample scale for test (more graph context is provided for prediction), results in better recommendations accuracy.
\end{itemize}

\begin{table}[t]
    \scriptsize
    \centering
	\setlength{\tabcolsep}{0.8mm}
    \begin{tabular}{|c|c|c|c|c|c|c|c|c|}
        \hline
        \multirow{2}{*}{Training $N$} & \multirow{2}{*}{Metric} & \multicolumn{7}{c|}{Number of Sub-graph Size $N$ When Testing}\\
        \cline{3-9}
        & & 20,000 & 30,000 & 50,000 & 70,000 & 90,000 & 110,000 & 130,000\\
        \hline
        \hline
        \multirow{1}{*}{30,000} & HR & 0.379 & 0.413 & 0.452 & 0.466 & 0.470 & 0.473 & 0.478\\
        \cline{2-9}
        \hline
        \multirow{1}{*}{50,000} & HR & 0.357 & 0.388 & 0.433 & 0.463 & 0.470 & 0.476 & 0.479\\
        \cline{2-9}
        \hline
        \multirow{1}{*}{70,000} & HR & 0.338 & 0.384 & 0.429 & 0.461 & 0.469 & 0.478 & 0.481\\
        \cline{2-9}
        \hline
        \hline
        \multicolumn{2}{|c|}{Testing Time (s)} & 70.3 & 94.1 & 148.8 & 207.8 & 261.6 & 309.3 & 351.7\\
        \hline
    \end{tabular}
    \vspace{-0.1in}
    \caption{Influence of the sub-graph sampling scale.}
    \label{tab:graphSampling}
    \vspace{-0.1in}
\end{table}

\vspace{-0.1in}
\subsection{Case Studies of \emph{\model}'s Explainability (RQ6)}
\begin{figure}
    \centering
    \includegraphics[width=\columnwidth]{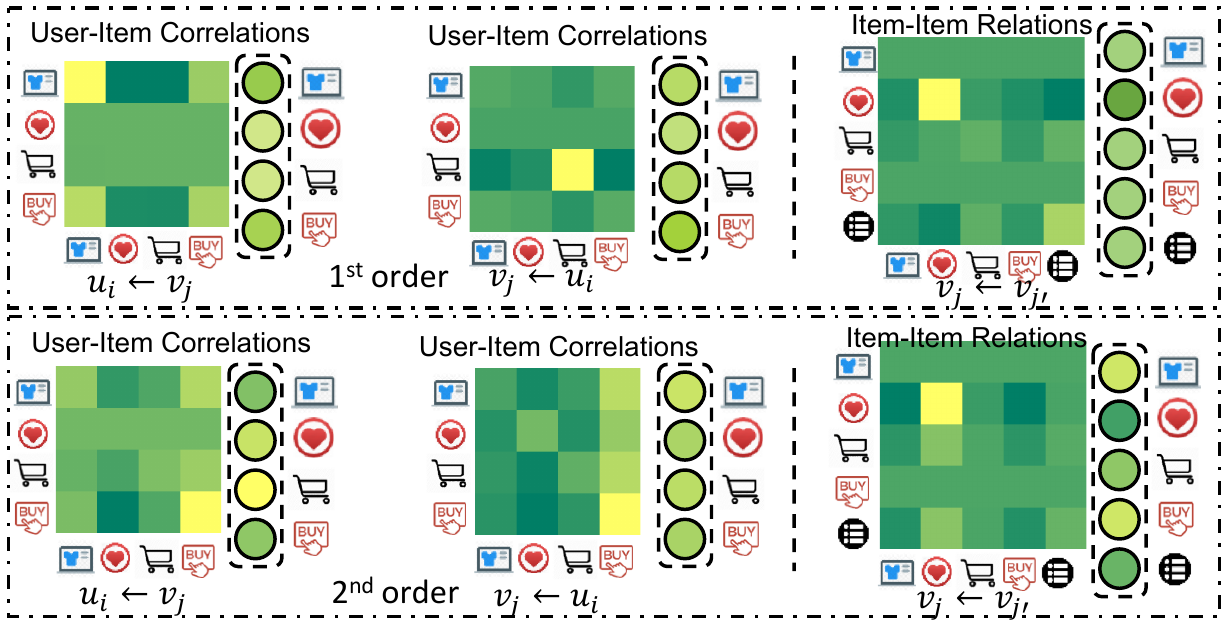}
	\vspace{-0.25 in}
    \caption{Visualized explicit relevance learned by \model.}
    \label{fig:att_caseStudy}
    \vspace{-0.15in}
\end{figure}

We visualize the learned explicit relevance scores of our \emph{\model} model for predicting purchases on retail data in Figure~\ref{fig:att_caseStudy}. We observe that different types of user-item interactions (4 types) and item-item relations (5 types) are correlated in a hierarchical and explainable manner (Brighter colors signal higher behavior relevance). In particular, Squares and circles represent the learned cross-type behavior dependencies in our type-wise behavior mutual relation encoder and cross-type behavioral pattern fusion, respectively. Both first- and second-order attentive weights are presented.


\section{Related Work}
\label{sec:relate}

\noindent \textbf{Recommendation with Multi-Relation Modeling}.
There exist many recommender systems which are developed to characterize user representations, with the consideration of different-typed relations from either user or item side~\cite{yu2018walkranker}. For example, to alleviate the data sparsity issue, a lot of social recommender systems have been proposed to boost the prediction performance via integrating the user-user social influential dependencies with the user-item interactive relations~\cite{socialaaai,fan2019graph}. Furthermore, another paradigm of multi-relation recommendation models leverage external knowledge graph information to construct different structural relations between a set of entities or items~\cite{wang2019knowledge,wang2019kgat}. Different from them, this work generalizes the joint modeling of multiplex collaborative relations and knowledge-aware item dependency in the multi-behavior recommendation.


\noindent \textbf{Graph Neural Network Recommender Systems}.
With the recent success of graph neural network in aggregating information from various relational data, many graph neural techniques have been proposed to learn user's preference over the graph-structured data for various recommendation tasks, such as GraphSAGE~\cite{hamilton2017inductive} and NGCF~\cite{wang2019neural} for encoding high-order user-item interactions, and GNNs for users' sequential behavior modeling~\cite{sessionaaai}. Inspired by the above research work, we propose a new method \model\ within the broader graph neural paradigm for multi-behavior recommendation, to solve its unique challenges resulting from relation heterogeneity between users and items.

\section{Conclusion}
\label{sec:conclusion}
This paper explicitly models type-specific user behavioral pattern in enhancing recommender systems. We devise a novel hierarchical graph transformer network, termed as \model, to perform the joint information aggregation over the user-item and item-item collaborative relations in multiple knowledge-aware behavior modalities. This refines type-specific behavior representations and encode their fine-grained interactive preference over items. Evaluation results on three datasets well validate our framework. Our future work is to fully deploy \model\ in an online working system to handing the streaming data in a recursive mode.

\section*{Acknowledgments}
We thank the anonymous reviewers for their constructive feedback and comments. This work is supported by National Nature Science Foundation of China (62072188, 61672241), Natural Science Foundation of Guangdong Province (2016A030308013), Science and Technology Program of Guangdong Province (2019A050510010). 

\bibliography{refs}

\end{document}